\documentclass[]{revtex4}

\usepackage[dvips]{graphicx}
\usepackage[dvips]{graphics}

\begin{document}

\begin{center}
{\LARGE \bf

Intra-Unit-Cell magnetic correlations near optimal doping in $\rm YBa_2Cu_3O_{6.85}$ }

\medbreak
{ \Large

L. Mangin-Thro$^a$, Y. Sidis$^a$, A. Wildes$^b$ and  P. Bourges$^{a,c}$
\\

 \medbreak 
}
\end{center}
 \large
$^a$ Laboratoire L\'eon Brillouin, CEA-CNRS, CEA-Saclay, 91191 Gif sur Yvette, France\\
$^b$ Institut Laue-Langevin, 71 avenue des martyrs, 38000 Grenoble, France\\
$^c$ email: philippe.bourges@cea.fr\\

 \bigbreak

{ \bf
Understanding high-temperature superconductivity requires a prior knowledge of the nature of the enigmatic pseudogap  metallic state \cite{Norman}, out of which the superconducting state condenses.  In addition to the electronic orders involving charge degrees of freedoms recently reported inside the pseudogap state \cite{Keimer,Davis13,Pepin}, a magnetic intra-unit-cell (IUC) order was discovered in various cuprates \cite{Fauque,Li-Nature,DeAlmeida,CC-review,CC-review2} to set in {\it just at} the pseudogap temperature, T*. In  nearly optimally doped YBa$_2$Cu$_3$O$_{6.85}$, polarized neutron scattering measurements, carried out on two different spectrometers, reveal new features. The order is made of finite size planar domains, hardly correlated along the c-axis. At high temperature, only the out-of-plane magnetic components correlate, revealing a strong Ising anistropy,  as originally predicted in the loop current model \cite{Varma06}. Below T*, a correlated in-plane response develops,  giving rise the  apparent tilt of the magnetic moment at low temperature \cite{Fauque,CC-review}. The discovery of these two regimes put stringent constraints  on the intrinsict nature of IUC order, tightly bound to the pseudogap physics.
}

 \bigbreak 

Polarized neutron scattering experiments  reported the existence of an IUC magnetic order in four different cuprates families \cite{Fauque,Li-Nature,DeAlmeida,CC-review,CC-review2,Mook,Li-PRB,ManginThro}. The IUC magnetic order that breaks time-reversal symmetry (TRS) but preserves the lattice translational symmetry (LTS). The IUC order develops below a temperature $\rm T_{mag}$ that matches the pseudogap temperature T* as defined by the resistivity measurement \cite{Ito,Ando} and where recent resonant ultrasound spectroscopy measurements in $\rm YBa_2Cu_3O_{6+x}$ show that the pseudogap state is a true symmetry breaking phase \cite{Shekhter}. While bulk neutron scattering measurements provides unambigous evidence for an IUC magnetic order, such an order was not detected by local magnetic probe measurements \cite{CC-review,Mounce,NMR}, a conundrum that could be understood in terms of finite size and slowly fluctuating magnetic domains \cite{Varma14}, whose existence has not been established so far. 

The observed IUC magnetism can simply be described as a q=0 antiferromagnetic state as it keeps the LTS,  but requires an  internal  staggered magnetic pattern inside each unit cell. Such a state is qualitatively consistent with  the loop current (LC) model for the pseudo-gap \cite{Varma06,Varma14}. While out-of-plane magnetic moments should be produced by the planar confinement of LC,  the observed magnetic moments also display an unexpected in-plane component, giving rise to a tilt of about $\sim$ 45$^{o}$ with respect to c-axis \cite{Fauque,Mook,Li-PRB,CC-review} at low temperature.   Several attempts to overcome this discrepancy were proposed: an additional spin response through spin-orbit coupling \cite{Varma-SO}, a geometrical tilt of moment through delocalisation of the LC over CuO octaedra or pyramids \cite{Weber,Lederer}, a superposition of LC states through quantum effects \cite{Varma-recent}. Besides the LC model, electronic phases breaking TRS, but preserving LTS, were proposed to preempt  either a  pair density wave state \cite{Agterberg} or a composite charge density wave state \cite{Wang}.  Alternative scenarii consider an IUC order based on a nematic-like state with either spin \cite{Fauque} or orbital \cite{Moskvin} moments on oxygens, or the entangled spin and orbital degrees of freedom within a model of magneto-electric multipoles \cite{Lovesey}. All of these theories give different interpretations of the origin of the tilt of the magnetic moment.

 A this stage, more experimental information are required on the size and lifetime of the magnetic domains and exact orientation of the magnetic moment to make progress on the understanding of the intrinc nature of the IUC magnetic order and its interplay with the pseudo-gap physics. This can be achieved around optimal doping. Indeed,  a fast decay of the magnetic intensity was reported from the underdoped to optimally doped samples observed in $\rm Bi_2Sr_2CaCu_2O_{8+\delta}$ compounds \cite{ManginThro}. Meanwhile, the magnetic intensity is proportional to $\rm T_{mag}$ in the underdoped regime, whereas this scaling suddenly breaks down upon approaching optimal doping. These effects could actually be triggered by a redistribution of the magnetic scattering in momentum space, namely a shortening of the magnetic correlation length upon increasing hole doping. In  nearly optimally doped $\rm YBa_2Cu_3O_{6.85}$ ($\rm T_c = 89K$, p = 0.15), these characteristic properties are present (see supplementary materials): the IUC magnetic order  still develops at a rather high temperature  $\rm T_{mag} \sim 200K$, but  the measured intensity at the Bragg position is  strongly reduced, as compared to more underdoped samples.

To gain a new  insight of the IUC magnetic order, we carried out on $\rm YBa_2Cu_3O_{6.85}$ a polarized neutron experiment using the multi-detector diffractometer D7, usually used to measure diffuse magnetic scattering. A survey of a few trajectories around different (1,0,L) $\bf{Q}$-positions reveals the persistence of magnetic scattering away from the Bragg reflections. In particular, for the trajectory across $\bf{Q}$ = (1,0,0.25), the contrast in intensities noticeably improves compared to the Bragg position $\bf{Q}$ = (1,0,0). In Fig.~\ref{Y685-Fig-1}.a-b, measurements around $\bf{Q}$ = (1,0,0.25) show that the magnetic signal increases upon cooling. At high temperature (300K) the magnetic signal is assumed to be flat whereas at low temperature (100K) there is a net enhancement of the magnetic intensity. From the H-dependences of Fig.~\ref{Y685-Fig-1}.b, one can extract the q-width (full width at half maximum) by a Gaussian fitting, $\rm \Delta{q}=0.025\pm0.004$ r.l.u. We observe no noticeable evolution of that q-width with temperature.  Actually, $\rm \Delta{q}$ is slightly broader than the resolution $\rm \Delta{q_{res}}=0.0195\pm0.001$ r.l.u. suggesting short-range correlations. One can extract a finite planar magnetic correlation length, $\rm \xi_{ab} = a/(\pi*\Delta{q_{i}})$ where $\rm \Delta{q_{i}}=\sqrt{\Delta{q}^{2}-\Delta{q_{res}}^{2}}$ is obtained by deconvolution from the resolution: $\rm \xi_{ab} \sim 20a \sim$ 75 \AA\  ($\rm a = 3.85\AA$ is the in-plane lattice parameter). Interestingly, finite magnetic correlation lengths generally imply the existence of a finite characteristic time scale. This may explain why the observation of the IUC magnetic order has not yet been corroborated by magnetic local probe measurements such as nuclear magnetic resonance\cite{Varma14}, due to their much longer time scales. In such a line of thought, the correlation length ought to stay finite even at lower doping to explain the absence of a magnetic signal in local probes. Finally, it is also worth pointing out that X-ray measurements recently revealed charge density order (CDW) \cite{Keimer,BlancoCanosa} which appears at lower temperature $\rm T_{CDW}<T_{mag}$ with correlation lengths $\rm \xi_{CDW} \sim 8a < \xi_{ab} \sim 20a$. That suggests the pre-eminence of the IUC order over the CDW instability.

Next, we average the intensity $I_i$ of N=10 detectors ($\overline{I}= \rm \sum\limits_{i=1}^N I_{i}$/N) around each $\bf{Q}$ = (1,0,L) position (corresponding to different rocking sample angles) to further improve the contrast of the magnetic intensity. The total magnetic intensity summed over 10 detectors is reported for two L values not at the Bragg position as a function of temperature (Fig.~\ref{Y685-Fig-2}.a). First, it reveals that the magnetic correlations start actually at a high temperature clearly above ${\rm T_{mag} \sim}$ 200K indicative of pre-transitional magnetic scattering. This is already depicted in Fig.~\ref{Y685-Fig-1}.a from H-scans. Second, the comparison of the total magnetic intensity at both L values (L=0.25 and L=0.5) in Fig.~\ref{Y685-Fig-2}.a shows that the magnetic intensities grow first similarly at both L and then more rapidly for L=0.25. That behavior suggests a change of the correlations along c* upon cooling with a gradual build-up as expected on approaching a phase transition. Using absolute units conversion, one can compare data from both 4F1 and D7 (Fig.~\ref{Y685-Fig-2}.b), from which one can estimate $\rm \Delta{q}=0.65\pm0.05$ r.l.u. corresponding to a very short correlation length along c*, $\xi_c \sim 0.5c$, at 100K. It is worth to emphasize that such broad L-dependence can explain the sharp decreasing of the peaked magnetic intensity versus doping, the L-integrated intensity being in constrast consistent with the ratio of both $\rm T_{mag}$. 

One can also use the temperature dependence of the averaged intensity in 10 detectors to estimate the directional components of the magnetic moment. In Fig.~\ref{Y685-Fig-2}.c-d, we report the temperature evolution of the 
the out-of-plane component $\rm M_{c}$  and the in-plane component $\rm M_{ab}$. Interestingly, both cross-sections exhibit a noticeable distinct temperature dependences. At high temperature, the in-plane component vanishes, meaning that the correlated magnetic fluctuations $<$M$^2$$>$ essentially correspond to the out-of-plane component. That strong Ising character (along c*) remains down to ${\rm T_{mag}}$. Below ${\rm T_{mag}}$, the in-plane component is increasing as well. As $\overline{I} \propto I_0/\xi_{ab}$ (where $I_0$ is the magnetic peak intensity), it indicates either that both magnetic components have a distinct reponse in temperature of their amplitude or that their correlation lengths exhibit a significantly different thermal evolution. Interestingly, the Ising character of the IUC order is masked at low temperature by the onset of the additional in-plane component. 

At 100K, one can estimate the ratio between $\rm M_{ab}$ and $\rm M_{c}$. That defines an apparent tilt angle, $\theta$ of the magnetic moment with respect to the c*-axis as $\rm tan(\theta)=M_{ab}/M_c$ \cite{CC-review}: from Fig.~\ref{Y685-Fig-2}.b-c, $\rm tan(\theta) \sim$ 0.84. That gives $\rm \theta \sim 40^{o}\pm 9^{o}$ for L $\sim$ 0 similar to what has been deduced previously for lower doping \cite{Fauque,Mook,CC-review}. That result dismisses all models where $\theta$ goes to zero at L=0 such as LC models that include apical oxygens as an explanation for the tilt \cite{Weber,Lederer}. The observed temperature dependence of $\theta$ is also not consistent with theories where the tilt direction is due to rigid geometrical factors. 

Going further, we applied the standard D7 polarization analysis relation for paramagnetic systems, typically valid for disordered magnetism\cite{Stewart} to get $\rm I_{mag}$ around H=0.88 and L=0 as a function of temperature (Fig.~\ref{Y685-Fig-3}.a). Interestingly, one observes a maximum in this quantity at around 200K corresponding to the transition temperature $\rm T_{mag}$ (see Fig.~S2.c-d measured on 4F1). At wave-vector (H,0,0) far away from H=1 (say $\rm H <0.8$), this cusp-like shape disappears. Such a shape is usually characteristic of critical scattering around the ordering temperature. However, here none of the magnetic cross-sections at H=0.88 displays any cusp as shown by the temperature dependence of the SF cross sections for two polarizations (Figs.~\ref{Y685-Fig-3}.b). This absence is actually not surprising as we observe finite correlation lengths. 

This implies a more subtle interpretation of the maximum in terms of the magnetic components, $<$M$^2_c$$>$ and $<$M$^2_{ab}$$>$, from which, one can rewrite $\rm I_{mag} \sim 2(<M_{ab}^{2}>-<M_{c}^{2}>)$. At 300K, both components are fluctuating. When $<$M$^2_c$$>$ starts to correlate at H=1 (Figs.~\ref{Y685-Fig-2}.c), the intensity at H=0.88 in the Z polarization decreases accordingly whereas $<$M$^2_{ab}$$>$ remains fluctuating (Figs.~\ref{Y685-Fig-3}.b), 
yielding an increase of $\rm I_{mag}$. Below 200K, $<$M$^2_{ab}$$>$ correlates as well and $\rm I_{mag}$ decreases accordingly. The cusp observation then further points toward a distinct behaviour for both magnetic components.

 \medbreak 

-------------------------------

Methods: 

We here report polarized neutron measurements on a $\rm YBa_2Cu_3O_{6.85}$ ($\rm T_c = 89K$) single crystal, previously used to study the spin dynamics \cite{Pailhes}, extending the study of the IUC order at a higher hole doping  p = 0.15 near optimal doping. The polarized neutron experiments have been performed on two spectrometers: the triple axis spectrometer 4F1 (Orph\'ee, Saclay) and the multi-detector diffractometer D7 (ILL, Grenoble) (see supplementary informations). For both, a polarizing super-mirror (bender) and a Mezei flipper are inserted on the incoming neutron beam in order to select neutrons with a given spin. In addition, a filter (pyrolytic graphite for 4F1 or beryllium for D7) is put before the bender to remove high harmonics. After the sample, the final polarization, ${\bf P}$, is analyzed by an Heusler analyzer on 4F1 whereas D7 is equipped with an array of polarizing benders in front of the detectors. For each wave vector $\bf{Q}$, the scattered intensity is measured in both spin-flip (SF) and non-spin-flip (NSF) channels. For the measurements on 4F1 the incident and final neutron wave vector are set to $\rm k_I = k_F = 2.57\AA^{-1}$. On the diffractometer D7, the incident wave-vector is taken as $\rm k_I = 1.29\AA^{-1}$. As we are using cold neutrons, we had about 4 times better Q-resolution compared to 4F1. Following previous studies \cite{Fauque,Mook,Li-Nature,Li-PRB,DeAlmeida,ManginThro}, the search for magnetic order in the pseudo-gap phase is performed on Bragg reflections $\bf{Q}$ = (1,0,L) with integer L values. The sample was then aligned in the scattering plane (1,0,0)/(0,0,1).

D7 is equipped with a fixed XYZ polarization mode\cite{Stewart} with Z being vertical and X and Y are pointing along arbitrary in-plane directions. For all of the three polarizations X, Y and Z, the SF and NSF channels were systematically measured over a few rocking angles of the sample in order to map the magnetic  scattering. We then performed data reduction adapting the standard procedure \cite{Stewart}  to force the magnetic scattering to be zero at 300K for any polarization direction. This is made in order to find evidence for small magnetic intensities (see supplementary information for details). The procedure includes background subtraction, flipping ratio, vanadium corrections. Further, following the procedure discussed above for 4F1, the conversion in absolute units has been performed using the Bragg peak $\bf{Q}$=(1,0,0) intensity.

 \medbreak 
-------------------------------

Acknowledgements: 

We wish to thank C.M. Varma, M.K. Chan and M. Greven for fruitful discussions. We also acknowledge financial support from the grant EXCELCIUS of the Labex PALM of the Universit\'e Paris-Saclay and  the project UNESCOS (contract ANR-14-CE05-0007) of the ANR.

\clearpage

\begin{figure}[t]
\includegraphics[width=12cm,angle=0]{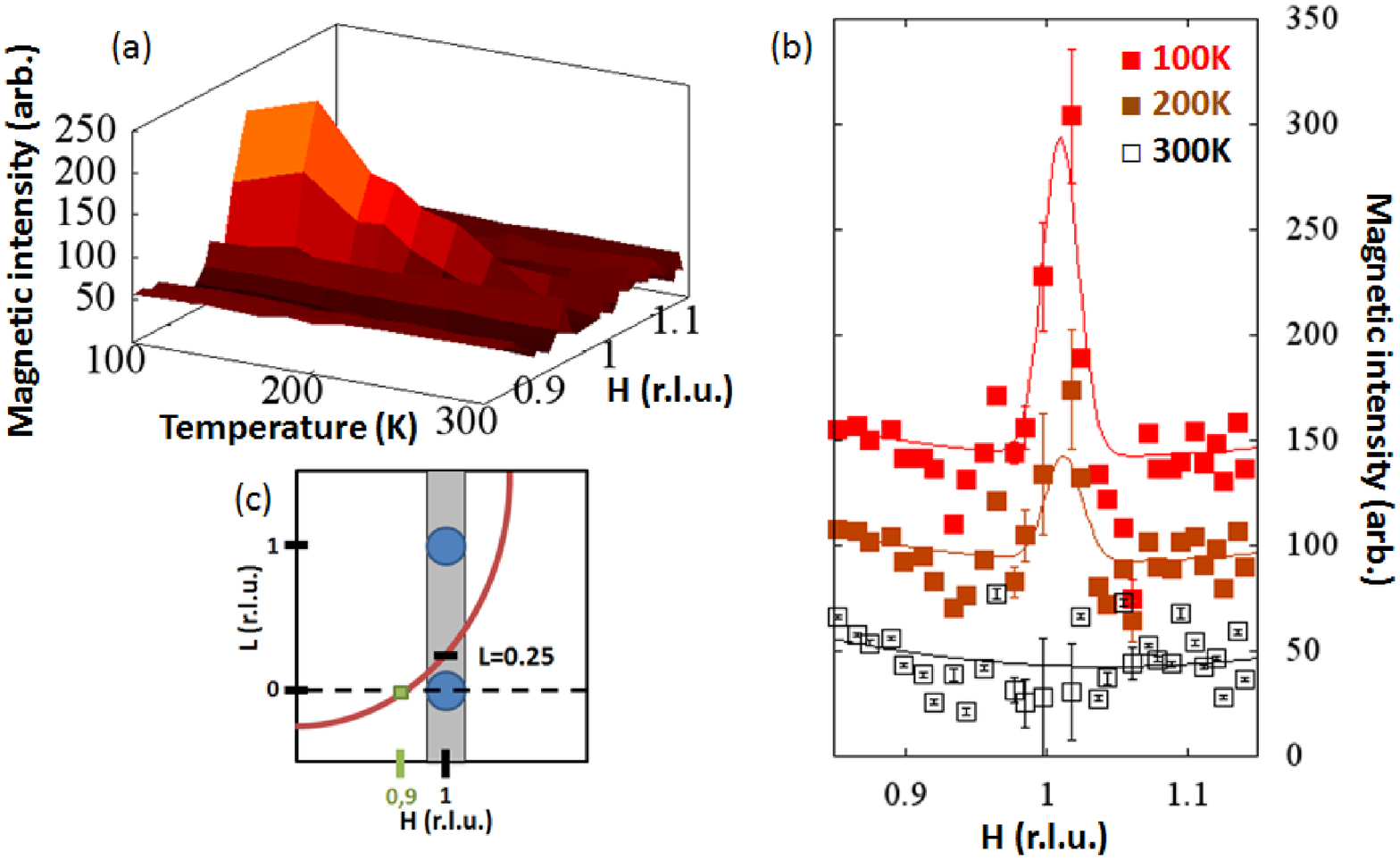}
\caption {
(color online) In-plane magnetic correlations. (a) Three dimensional plot of the maximum of the magnetic intensity in the SF channel around $\bf{Q}$ = (1,0,0.25) as a function of temperature and wave-vector. (b) Cuts of the 3D-map at three temperatures: 100K (red squares), 200K (brown squares) and 300K (black empty squares). Data obtained in channels X and Z are here averaged. c)    schematic view of the reciprocal space probed on D7 with in red the measured trajectory going through (1,0,0.25).  r.l.u. stands for reduced lattice units. }
\label{Y685-Fig-1}
\end{figure}

\begin{figure}[t]
\includegraphics[width=12cm,angle=0]{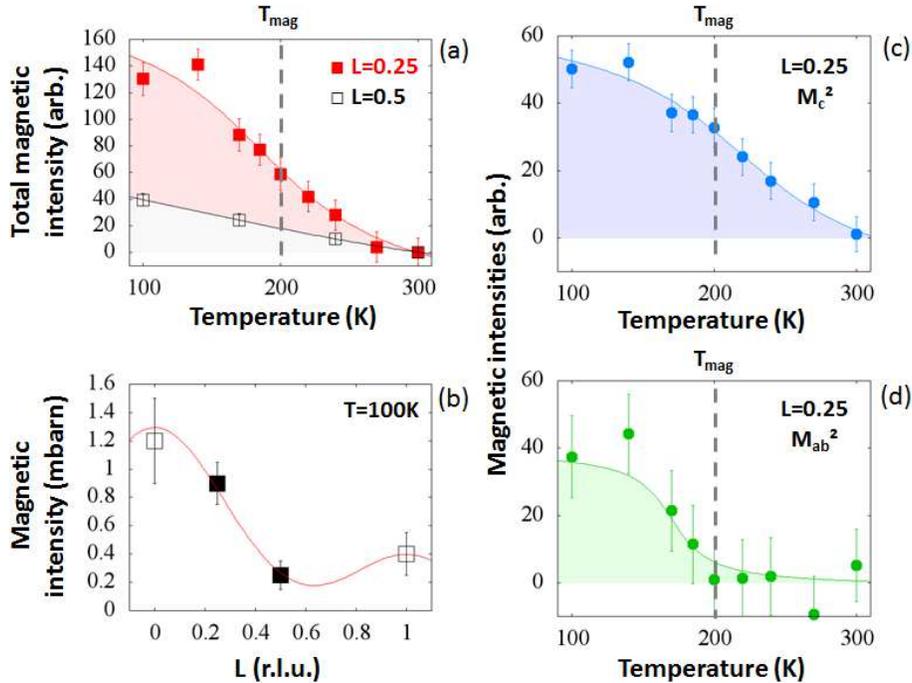}
\caption {
(color online) (a) Temperature dependences of the total magnetic intensity, $\rm \overline{I}^{SF}_{X}+\overline{I}^{SF}_{Y}+\overline{I}^{SF}_{Z}$ (background subtracted) for two L. (b) Magnetic intensity in absolute units from 4F1 (empty symbols) and D7 (full symbols) versus L at 100K. Temperature dependence of the magnetic components for L=0.25: (c) the out-of-plane component $\rm M_{c}^{2}$ and (d)  the in-plane component $\rm M_{ab}^{2}$ (see supplementary information).  Solid lines are guides to the eye. }
\label{Y685-Fig-2}
\end{figure}

\begin{figure}[t]
\includegraphics[width=12cm,angle=0]{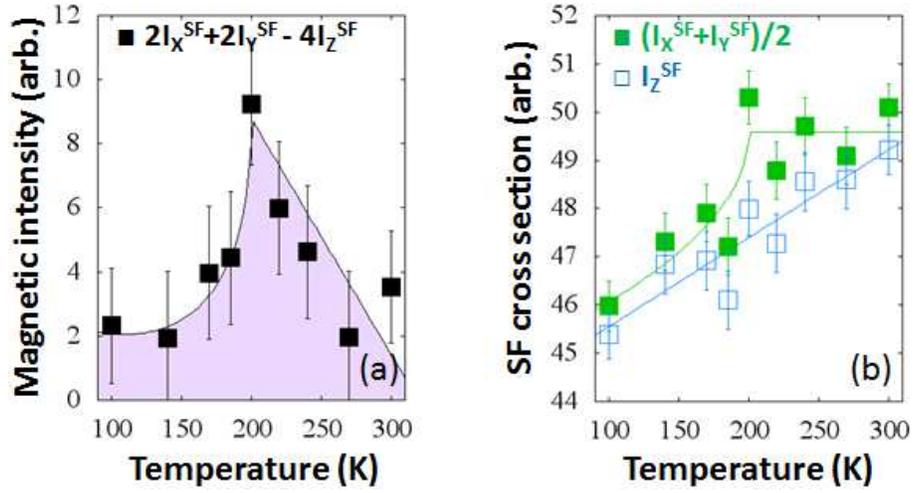}
\caption {
(color online) (a) Polarization analysis around $\bf{Q}$ $\sim$ (0.88,0,0) of $\rm I_{mag}=(2I^{SF}_{X}+2I^{SF}_{Y}-4I^{SF}_{Z})$. The maximum of intensity, which is only for $\bf{Q}$ $\sim$ (0.88,0,0), appears around $\rm T_{mag} \sim 200K$. (b) Temperature dependences of the SF cross section in the $\rm (I^{SF}_{X}+2I^{SF}_{Y})/2$ (green squares) and $\rm I^{SF}_{Z}$ (blue squares) polarizations for H=0.88. $\rm I^{SF}_{Z}$ typically probes the magnetic fluctuations along c, $<$M$^2_c$$>$ whereas $\rm (I^{SF}_{X}+2I^{SF}_{Y})/2$ is proportional to $<$M$^2_{ab}$$>$. The background is decreasing with temperature as expected for a spin incoherent background scattering. Solid lines are guides to the eye.
}
\label{Y685-Fig-3}
\end{figure}

\clearpage

\title{Supplementary information}

\maketitle

We here give additional information necessary for the analysis of the polarized neutron data. We report polarized neutron measurements on a $\rm YBa_2Cu_3O_{6.85}$ ($\rm T_c = 89K$) single crystal of mass $\sim 9.5$ g. The polarized neutron experiments have been performed on two spectrometers: the triple axis spectrometer 4F1 ( Laboratoire L\'eon Brillouin, Orph\'ee, Saclay) and the multi-detector diffractometer D7 (Institut Laue Langevin, Grenoble).
Many details concerning polarized neutron and flipping ratio on 4F1 has been already discussed in Refs. \cite{ManginThro,Fauque,Mook,Baledent-YBCO,Li-Nature,Li-PRB,DeAlmeida,CC-review,CC-review2} in the context of the magnetic intra unit cell in high temperature superconducting cuprates. 

For the measurements on 4F1, the incident and final neutron wave vector are set to $\rm k_I = k_F = 2.57\AA^{-1}$ and the energy resolution is $\sim$ 1 meV. On D7, the incident wavelength is taken as $\rm \lambda_I = 4.86\AA$ ($\rm \equiv k_I = 1.29\AA^{-1}$), which corresponds to $\rm E_I = 3.47 meV$. Since it is a diffractometer, neutrons with different final energies around $\rm E_I$ are measured. That corresponds to an integration of the magnetic fluctuations in an energy range between $\rm -k_{B}T$ to $\rm E_I$, where $\rm k_{B}$ is the Boltzmann constant.

\begin{figure}[h]
\renewcommand*{\thefigure}{S\arabic{figure}}
\includegraphics[width=9cm,angle=0]{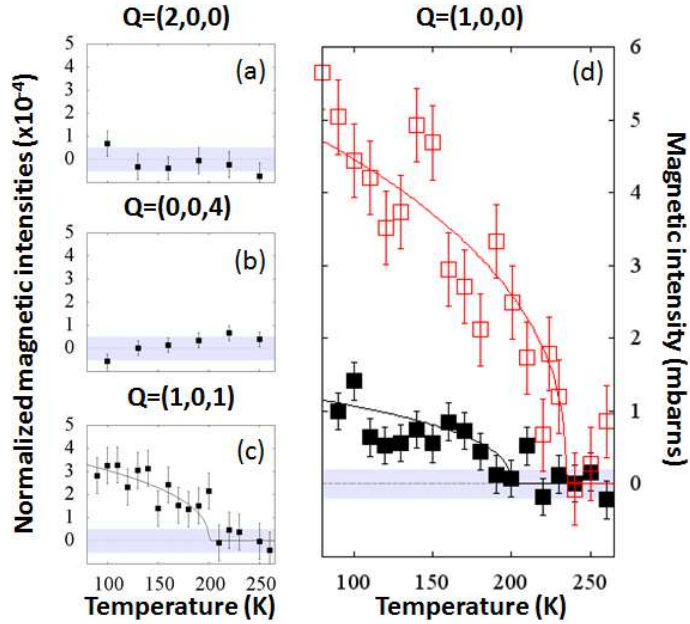}
\caption{
(color online) 4F1 data:  Temperature dependencies of normalized magnetic intensity (=${\rm I_{IUC}/I_{NSF}}$ where $\rm I_{IUC}$ is the expected magnetic intensity of the IUC order) measured at the wave vector (a) $\bf{Q}$ = (2,0,0), (b) $\bf{Q}$ = (0,0,4) as references and at (c)-(d) $\bf{Q}$ = (1,0,L) for the IUC magnetic order study in the $\rm \bf{P} \parallel \bf{Q}$ configuration. (c) L = 1, (d) L = 0. The data for $\bf{Q}$ = (1,0,0) ($\rm YBa_2Cu_3O_{6.85}$, full squares) have been calibrated in absolute units using the intensity of the Bragg peak $\bf{Q}$ = (0,0,4) \cite{Fauque} and further compared to the $\rm YBa_2Cu_3O_{6.6}$ study (red empty squares)\cite{Mook}.
}
\label{Y685-Fig-S1}
\end{figure}

\section*{\label{fr} Flipping ratio}

The intra-unit-cell (IUC) magnetic order produces a signal at the same position as the nuclear Bragg reflections \cite{CC-review,CC-review2}. The magnetic signal is much weaker than the nuclear signal but it acts to flip the neutron spin. As polarized neutron measurements enable to distinguish nuclear and magnetic scattering, they are a very elegant  tool to observe the IUC magnetic order. The scattered intensity at a given wave vector $\bf{Q}$ is then measured in both spin-flip (SF) and non-spin-flip (NSF) channels. The flipping ratio, $\rm FR=I_{NSF}/I_{SF}$, defines the quality of the polarization. A typical flipping ratio is of the order of 40 for 4F1. Furthermore, polarized neutron scattering allows to separate spin orientation along three orthogonal directions. For the configuration where the polarization $\bf{P}$ is parallel to $\bf{Q}$, where the magnetic scattering is exclusively in the SF channel \cite{CC-review}. Consequently, we essentially carried out measurements with that polarization on 4F1. 

We show on Fig.~\ref{Y685-Fig-S1} the results from the spectrometer 4F1 as a function of the temperature in both SF and NSF channels, for the neutron polarization $\rm \bf{P} // \bf{Q}$. To separate the magnetic scattering, it is convenient to estimate the inverse of the flipping ratio, which is given by:

\begin{equation}
1/FR(T)=I_{SF}/I_{NSF}=1/FR^0(T)+I_{IUC}/I_{NSF}
\label{ISF}
\end{equation}

$\rm 1/FR^{0}$ represents the polarization leakage (from the NSF channel into the SF channel), characterizing the neutron beam polarization quality of the instrument. It is defined at the highest measured temperature. 1/FR-$\rm 1/FR^0$ (=${\rm I_{IUC}/I_{NSF}}$ where $\rm I_{IUC}$ is the expected magnetic intensity for the IUC order) represents the normalized magnetic intensity reported in Fig.~\ref{Y685-Fig-S1}. This quantity is used in order to easily compare different studies of the magnetic intensity on different samples. At the first approximation, $\rm 1/FR^0$ is supposed to be temperature independent.  However, it appears empirically to be slightly temperature dependent \cite{Baledent-YBCO}. This T-dependence is determined by measurements at Bragg peaks where the magnetic signal can be ignored such as at large $|\bf{Q}|$ where any magnetic signal becomes vanishingly small. In Fig.~\ref{Y685-Fig-S1}.a-b, the normalized magnetic intensity  obtained at $\bf{Q}$ = (2,0,0) and $\bf{Q}$ = (0,0,4) is reported where a magnetic signal (if any) is beyond a threshold of detection ($<$ 5x10$^{-5}$). That defines in our experiment, $\rm 1/FR^{0}(T)$ as the average of the inverse of the flipping ratio at $\bf{Q}$ = (2,0,0) and $\bf{Q}$ = (0,0,4) shown in Fig.~\ref{Y685-Fig-S1}.a-b where no magnetic signal is observed, signifying that those two wave-vectors are good reference positions for corrections. 

\begin{figure}[h]
\renewcommand*{\thefigure}{S\arabic{figure}}
\includegraphics[width=9 cm,angle=0]{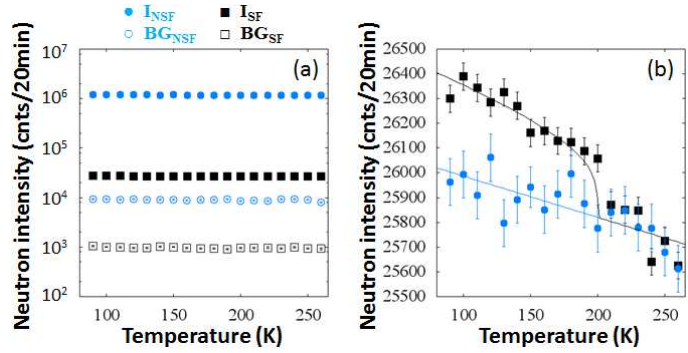}
\caption {
(color online) (a) Temperature dependencies of the raw neutron scattering intensities measured at the Bragg peak $\rm \bf{Q}$ = (1,0,1) (full symbols) and at $\rm \bf{Q}$ = (0.9,0,1) (empty symbols), the background position, in the NSF channel (blue) and in the SF channel (black). (b) Temperature dependencies of intrinsic Bragg scattering (background subtracted) at $\rm \bf{Q}$ = (1,0,1) in the SF channel (black) and in the NSF channel (blue) divided by a T-dependent bare flipping ratio $\rm FR^{o}(T)$, so that at high temperature there is no magnetic signal. Both pictures shows measurements in the $\bf{P} \parallel \bf{Q}$ configuration.
}
\label{Y685-Fig-S2}
\end{figure}

In order to extract the magnetic intensity, we  measured the Bragg reflections $\bf{Q}$ = (1,0,L) (L either 0 or 1) as a function of the temperature. As the Bragg intensity appeared on top of a background, we also needed to measure the intensity at $\bf{Q}$ = (0.9,0,L) to get the temperature dependence of the background in both SF and NSF channels. Fig.~\ref{Y685-Fig-S2}.a shows the temperature dependence of the SF and NSF scattering for $\rm \bf{Q}$ = (1,0,1) and $\rm \bf{Q}$ = (0.9,0,1). Compared to Bragg reflection, $\bf{Q}$ = (0.9,0,L) is small. That background is then subtracted from the signal at $\bf{Q}$ = (1,0,L) to keep only the intrinsic temperature dependence of the Bragg peak. Fig.~\ref{Y685-Fig-S1}.b shows the temperature dependence of the SF and NSF scattering normalized at the highest measured temperature $\sim$ 260K after the background has been removed. The NSF signal has been further divided by the T-dependent $\rm 1/FR^0(T)$. One can observe an enhancement in the SF channel corresponding to the appearance of the magnetic signal below $\sim$ 200K on top of NSF signal normalized by $\rm 1/FR^0(T)$.

Fig.~\ref{Y685-Fig-S1}.c shows normalized magnetic intensity for the Bragg reflections $\bf{Q}$ = (1,0,1). Starting from zero scattering at high temperature, a magnetic signal appears below $\rm T_{mag} \sim 200K\pm20K$. Our measurements first made on the Bragg reflection $\bf{Q}$ = (1,0,1) are confirmed by those on $\bf{Q}$ = (1,0,0) (Fig.~\ref{Y685-Fig-S1}.d) where the data are further compared to a previous study on $\rm YBa_2Cu_3O_{6.6}$\cite{Mook}. Such a value for $\rm T_{mag}$ agrees with recent accurate determination of T* from resistivity \cite{alloul-EPL}. The magnetic signal displays a characteristic T dependence that can be fitted by $\rm (1 - T/T_{mag})^{2\beta}$ with $\rm 2\beta = 0.4 \pm 0.17$ a value similar to what was found in $\rm YBa_2Cu_3O_{6.6}$\cite{Mook}. 

Once $\rm 1/FR-1/FR^o = I_{IUC}/I_{NSF}$ is determined one can convert $\rm I_{IUC}$ in absolute units following a method described in previous studies\cite{Fauque,CC-review}. For $\bf{Q}$ = (0,0,4), the square of the structure factor corresponds to $\rm \sim 7$ barns \cite{Fauque}. All cross-sections are given per formula unit (f.u.). Then, to get the Bragg intensity, we have applied the following conversion factor: $\rm [I_{NSF}(1,0,L)*7/I_{NSF}(0,0,4)]$. In our present case, $\rm I_{NSF}(1,0,0)/I_{NSF}(0,0,4)\sim4/7$, so that, if the amplitude of the magnetic intensity for both $\bf{Q}$ = (1,0,L) is $\rm \sim 3*10^{-4}$ of the nuclear intensity, for $\bf{Q}$ = (1,0,0) the magnetic intensity is of 1.2 mbarn/f.u.. Applying the same procedure for $\bf{Q}$ = (1,0,1), one gets $\rm \sim 3$ times weaker SF intensity, 0.4 mbarns/f.u.. Compared to previous measurements performed on $\rm YBa_2Cu_3O_{6.6}$ at $\bf{Q}$ = (1,0,0) \cite{Mook} and $\bf{Q}$ = (1,0,1) \cite{Fauque}, the magnetic intensity in $\rm YBa_2Cu_3O_{6.85}$ is $\rm \sim 4$ times weaker (Fig.~\ref{Y685-Fig-S1}.d). Following the same hypotheses as before to determine the magnetic moment\cite{CC-review} and keeping in mind that the magnetic intensity is proportional to $\rm M^2$, the magnetic moment is here estimated to be only $\rm \sim 0.05\mu_{B}$ compared  to $\rm 0.1\mu_{B}$ for under-doped YBCO \cite{Fauque,CC-review}.

\section*{\label{d7} D7 procedure}

The current set-up of the cold neutron diffractometer D7 (ILL, Grenoble) has been given in refs. \cite{Stewart,Ehlers}. It makes use of a triple-blade monochromator which selects the incoming neutrons. A beryllium filter after the monochromator removes high harmonics. A supermirror polarizer and the Mezei flipper select neutrons with a given spin. D7 is equipped with a fixed XYZ polarization mode at the sample position. Small guiding magnetic fields are produced from a set of electric coils. The final polarization is then analysed by 66 supermirror benders over a wide range of scattering angles. Two detectors are located after each bender, giving a variable flipping ratio for each detector ranging from 15 to 35. 

Both 4F1 and D7 allow the polarization to be arbitrarily oriented. On 4F1, as already described in previous papers \cite{Fauque,Mook,Baledent-YBCO,Li-Nature,Li-PRB,DeAlmeida,ManginThro,CC-review,CC-review2}, the polarization directions can be adapted to any direction relative to $\bf{Q}$. Contrary to the 4F1 set-up, the directions of the polarization on D7 are along arbitrary directions unrelated to $\bf{Q}$. The polarizations X and Y are in the scattering plane whereas Z is perpendicular to the plane thus always perpendicular to $\bf{Q}$. The direction X on D7 is along a direction making an angle 24$^{o}$ from $\bf k_{i}$ \cite{Ehlers}. Measurements were performed in both SF and NSF channels, for the three X, Y, Z polarizations. We used the Lamp software provided by ILL to analyse the data (see http://www.ill.eu/instruments-support/computing-for-science/cs-software/all-software/lamp/). We applied the standard procedure which consists in a background subtraction, a flipping ratio correction and a vanadium normalization to get intensities in absolute units for each channels. 

We carried out measurements of the background coming from an empty sample holder so that we could compare measurements with and without the sample. A quartz rod of the size of our $\rm YBa_2Cu_3O_{6.85}$ sample was put in the cryostat in order to get the flipping ratio correction. The diffuse scattering from amorphous quartz is entirely nuclear. Consequently, it indicates the ratio of the correctly attributed non-spin-flip scattering cross section to the wrongly attributed spin-flip scattering cross section over all the detectors (132 detectors). This gives the flipping ratio FR for each detector. In the limit of large FR \cite{Stewart}:

\begin{equation}
I_{SF}^{corr}=I_{SF}-\frac{1}{FR-1}(I_{NSF}-I_{SF})
\label{ISFcorr}
\end{equation}
\begin{equation}
I_{NSF}^{corr}=I_{NSF}+\frac{1}{FR-1}(I_{NSF}-I_{SF})
\label{INSFcorr}
\end{equation}

We apply the flipping ratio correction (as Eqs. \ref{ISFcorr} and \ref{INSFcorr}) on our dataset in YBCO using flipping ratios determined using the quartz sample. Actually, that correction is insufficient. Indeed, the amorphous quartz sample gives broad diffuse scattering and the flipping ratios are generally reliable for correcting diffuse scattering. However, systematic errors can occur when measuring different scattering, such as an intense Bragg peak which naturally is better collimated and samples a different weighted average of the integrated polarization efficiency of the instrument. As a result, we actually observed larger flipping ratio on our YBCO single crystal than on the quartz reference. To improve the analysis, we made the assumption that no magnetic signal is present at 300K. We then  adjusted the flipping ratio for each detector so that the scan is featureless at 300K for each $\bf{Q}$=(1,0,L) and for each polarization. We then applied the flipping ratio correction of Eq. \ref{ISFcorr} with a corrected set of flipping ratios. The intrinsic magnetic signal appears then clearly at 100K on top of the flat scan at 300K (Fig. 2b). Finally, we performed measurements on a vanadium rod, to correct the data for detector efficiency and to be able to calculate the cross section in absolute units. Further, following the procedure discussed above for 4F1, the conversion in absolute units has been cross-checked using the Bragg peak $\bf{Q}$=(1,0,0) intensity. 

\section*{\label{cross-sections} Neutron cross-sections}

For the case of an ordered state with a magnetic moment pointing along a generic direction $\rm M=(M_a,M_b,M_c)$, one can estimate the  magnetic neutron cross-sections for each neutron polarization $\rm I_{\alpha}$ with $\rm \alpha=(X,Y,Z)$. For a twinned sample, both directions (1,0) and (0,1) are equivalent and one can define $\rm M_{ab}=\sqrt{M_a^2+M_b^2}$ as the in-plane component and $\rm M_{c}$ as the out-of-plane component\cite{CC-review}. The neutron intensity for any arbitrary direction of the polarization $\rm \alpha=(X,Y,Z)$ can be written as \cite{Stewart}, $\rm (\frac{d\sigma}{d\Omega})_{sf}^{\alpha}=I^{SF}_{\alpha}+ BG_{\alpha}$ where $BG_{\alpha}$ is the background. For a XYZ-polarization turned by an arbitrary angle $\rm \beta$ between X and $\bf{Q}$, $\rm I^{SF}_{\alpha}$ can be written as,

\begin{equation}
I^{SF}_{Z} \propto (1-q_l^2)M_{c}^{2}+\frac{q_l^2}{2}M_{ab}^{2} \sim M_{c}^{2}
\label{IZ}
\end{equation}
\begin{equation}
I^{SF}_{Y} \propto \frac{1}{2}M_{ab}^{2}+ I_Z ~ sin^2 \beta
\label{IY}
\end{equation}
\begin{equation}
I^{SF}_{X} \propto \frac{1}{2}M_{ab}^{2}+ I_Z ~ cos^2 \beta
\label{IX}
\end{equation}

where $\rm q_l= \frac{2 \pi}{c} L /|{\bf Q}|$ tends to zero in our study for L$\sim$ 0. The same formulas can be applied for the 4F1 setup but with $\beta=0$, the polarization along X for 4F1 being along $\bf{Q}$. 

 The total magnetic intensity is obtained from the sum of the three cross-section in the SF channel 
$\rm \overline{I}^{SF}_{X}+\overline{I}^{SF}_{Y}+\overline{I}^{SF}_{Z}$ whose temperature dependences is reported in Fig.2 of the manuscript for two L. One sees that $\rm I^{SF}_{X}+I^{SF}_{Y}-I^{SF}_{Z}  \propto  M_{ab}^{2}$. In the limit  $\rm q_l \to 0$ (as it is the case for L=0.25), $I_{Z} \propto M_{c}^{2}$. That enables us to study both components of the magnetic moment $\rm \bf{M}$ separately as done in the manuscript. The cross-section in the SF channel $\rm \overline{I}^{SF}_{Z}$ essentially corresponds to the out-of-plane magnetic component, $\rm M_{c}^{2}$, and conversely the quantity $\rm \overline{I}^{SF}_{X}+\overline{I}^{SF}_{Y}-\overline{I}^{SF}_{Z}$ directly probes the in-plane component $\rm M_{ab}^{2}$. 
Following the definition of the magnetic moment, the tilt angle $\theta$, schematically drawn in Fig.~3.b, is defined by the angle of the magnetic moment with respect to the c axis. It is related to magnetic components as $\rm tan(\theta) = \frac{M_{ab}}{M_{c}}$.

In the opposite limit of disordered magnetism, one generally considers the cross section for isotropic magnetic correlations, {\it i.e.} $\rm M_a^2=M_b^2=M_c^2$. This corresponds to a model for paramagnetism that is extensively applied on D7 \cite{Stewart}. $\rm I_{mag}$ is written as,

\begin{equation}
I_{mag}=2I^{SF}_{X}+2I^{SF}_{Y}-4I^{SF}_{Z}
\label{paramag}
\end{equation}

The standard D7 full XYZ polarization analysis separates the (para)magnetic intensity independently from the background (such as nuclear and spin-incoherent scattering). The only assumption is that the background is not polarization dependent. This is readily the case at position away from the Bragg peaks. It is worth to note that Eq. \ref{paramag} determines the total magnetic scattering only for paramagnetic scattering. In this limit, it gives   $\rm I_{para}={2\over3} g^2 S(S+1)$ where g is the Land\'e factor and S is the effective spin. 
This does not apply for anisotropic magnetic systems. 

\section*{\label{average} Averaging intensities}

In order to get better contrast of the intensities for particular wave-vectors, we averaged intensities $\rm I_i$ from each N detectors using $\rm \overline{I}= \sum\limits_{i=1}^N I_{i}$/N. That quantity is actually related to the integrated area of the peak. The intensity at each detector $\rm I_i$ ({\it i.e.} at a wave vector $q_i$) is a sum of a background $BG_{i}$ and a contribution coming from a peak centered at $q_0$ with a FWHM of $\Delta q$ and a magnetic amplitude  $\rm I_{0}$. For a Gaussian description of the peak, $\rm I_i$ is written as 

\begin{equation}
I_{i} = BG_{i} + I_{0} \exp(-4 \ln 2(q_i-q_0)^{2}/\Delta q^{2}) 
\label{Ii}
\end{equation}

The averaged intensity $\rm \overline{I}$ can be written as,
 
\begin{equation}
\overline{I}=\sum\limits_{i=1}^N I_{i}/N = \overline{BG} + I_{0} \frac{\int_{q_1}^{q_N} \exp(-4 \ln 2(x-x_0)^{2}/\Delta q^{2})}{\int _{q_1}^{q_N} dq } = \overline{BG} + I_{0} \frac{\Delta q}{q_N-q_1} \sqrt{\frac{\pi}{4 \ln 2}}
\label{sum}
\end{equation}

$\rm (q_{N}-q_{1})$ is the width on which we are averaging intensities, $\rm  \overline{BG}$ is the averaged background level taken at high temperature near the peak. The quantity plotted in Fig. 3 of the manuscript is  $\rm \overline{I} - \overline{BG}$, it is proportional to the product of the intensity at the maximum and of the q-width: $\rm \propto I_{0} \Delta q$. As $\Delta q \propto 1/\xi_{ab}$, $\overline{I}$ then typically measures the ratio of the magnetic intenisity to the in-plane correlation length: $\overline{I} \propto I_{0}/\xi_{ab}$.

\end{document}